\documentclass{moriond}
\usepackage{ptdr-definitions}
\usepackage{hepunits}
\usepackage{heppennames2}
\usepackage{lineno}

\def\be{\begin{equation}}
\def\ee{\end{equation}}
\def\bea{\begin{eqnarray}}
\def\eea{\end{eqnarray}}

\newcommand{\Photo}{}

\begin{document}

\vspace*{4cm}
\title{Flow harmonics in heavy ion physics at CMS and ATLAS}

\author{\href{https://gkrintir.web.cern.ch/}{Georgios K. Krintiras} (on behalf of the CMS and ATLAS Collaborations)}

\address{The University of Kansas Department of Physics and Astronomy, 1251 Wescoe Hall Dr, Lawrence, US}

\maketitle\abstracts{
How can we gain a detailed insight into the hydrodynamic response of the system created in heavy ion collisions to the fluctuating initial geometry and viscous effects? Do we create a strongly interacting medium in proton-nucleus and proton-proton collisions, or rather a system of partons undergoing few scatterings? To what extent can we discriminate between initial momentum correlations and flow generated as a response to the initial geometry via interactions in the final state? Do measurements of identified particle flow confirm the observations from inclusive charged hadrons? An experimental overview of anisotropic flow measurements, ranging from large down to the smallest collision systems, is given in these proceedings. 
}
\copyright
\section{Introduction}
\label{sec:intro}

One of the main purposes of the Large Hadron Collider (LHC) is to create a hot and dense, strongly interacting QCD medium, referred to as the quark-gluon plasma (QGP).
The study of ultrarelativistic heavy ion collisions, including the decomposition of azimuthal particle distributions into Fourier series, revealed QGP properties consistent with a collectively expanding (``flowing'') medium. The associated flow vectors $V_n \equiv v_ne^{in\Psi_n}$, where $v_n$ is the magnitude of the $n^{\text{th}}$-order Fourier harmonic and $\Psi_n$ its phase (also known as the $n^{\text{th}}$-or order ``symmetry plane angle''), reflect the hydrodynamic response of the medium to the transverse overlap region and its subnucleon fluctuations. Measurements of flow vectors, their event-by-event fluctuations and correlations between different orders of harmonics or symmetry planes provide input to QGP modeling, in particular, details about initial-state conditions and the dynamics of the subsequent deconfined phase~\cite{Gajdosova:2020nvb}.

In ``small'' collision systems, features similar to those observed in heavy ion collisions are revealed when the same observables are used in conjunction with high particle multiplicities. It thus remains imperative that LHC experiments, like ATLAS~\cite{Aad:2008zzm} and CMS~\cite{Chatrchyan:2008aa}, pursue their quest for a medium of a similar origin as in measurements involving heavy ion collisions. It is not yet clear, however, to what extent is the $V_n$ driven by the initial spatial anisotropy and/or whether competing mechanisms, \eg, gluon field momentum correlations at the initial state, contribute to the seen final-state anisotropy. In all cases, utmost caution must be exercised when interpreting flow-related signatures, specially in small systems, given the potential contamination from nonflow effects.

\section{Flow in heavy ion collisions}
\label{sec:AA}

Although $v_n$ measurements had led to ``the  discovery  of the  perfect  liquid''~\cite{Rafelski:2019twp} and contributed significantly to the understanding of the QGP, e.g., Refs.~\cite{Aaboud:2018ves,Sirunyan:2017pan}, details of the initial-state conditions and the subsequent dynamics can be probed by more complex observables: event-by-event flow fluctuations, decorrelations of flow vectors, higher order $V_n$, in particular their linear and nonlinear components, and correlations between different orders of  $v_n$ or $\Psi_n$.

\begin{figure}[!htb]
\centering
\begin{minipage}{0.5\textwidth}
\centering
\includegraphics[width=0.75\textwidth]{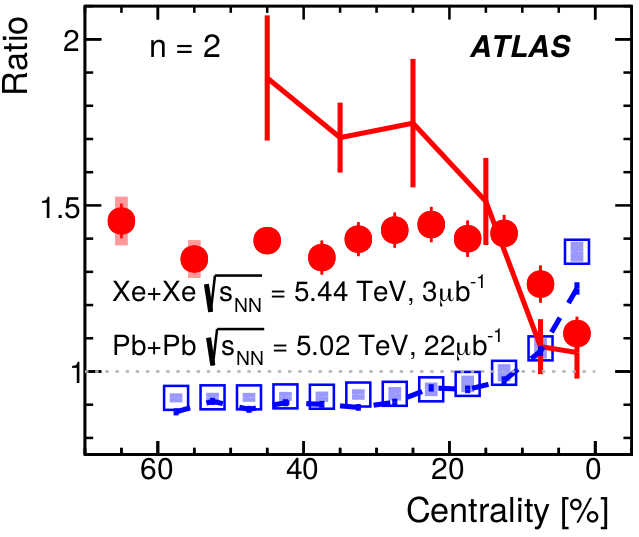}
\end{minipage}\hfill
\begin{minipage}{0.5\textwidth}
\centering
\includegraphics[width=0.99\textwidth]{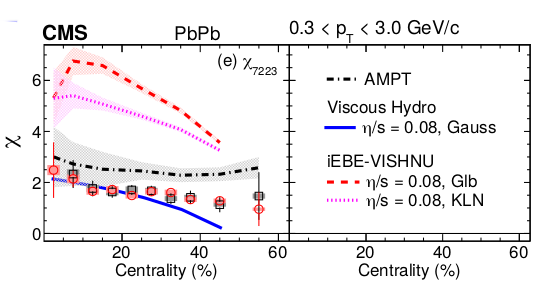}
\end{minipage}
\caption{Left: Ratios of flow decorrelations and $v_2$ from PbPb and XeXe collisions at 5.02 and 5.44\,\TeVns, respectively~\protect\cite{Aad:2020gfz}, compared to hydrodynamic models. Right: The nonlinear  response  coefficient $\chi_{7223}$ at 2.76 and 5.02\,\TeVns, as a function of centrality~\protect\cite{Sirunyan:2019izh}, compared to hydrodynamic models with different initial-state conditions.}
\label{fig:fig1}
\end{figure}

Investigations of such fluctuations can be conducted by measuring the \pt and centrality dependence of multiparticle azimuthal cumulants. First measurements~\cite{Aaboud:2019sma} of the four-particle cumulants $v_2\{4\}$ and $v_3\{4\}$ revealed that flow fluctuations are non-Gaussian, while $v_4\{4\}$ indicated that flow fluctuations are affected by the so-called volume or centrality fluctuations within a fixed centrality bin. Flow vector decorrelations can be studied by forming a factorization ratio $r_n (\pt, \eta)$ such that when $v_n$ and/or $\Psi_n$ decorrelate the $r_n$ deviates from unity. These measurements can provide important constraints on the longitudinal structure, currently a challenge for three-dimensional hydrodynamic models. More specifically, comparison of $r_n$ in PbPb and XeXe collisions~\cite{Aad:2020gfz} revealed that models tuned to describe the $v_n$ in both systems failed to reproduce the $r_n$, as it can be seen in Fig.~\ref{fig:fig1} (left). Higher order $V_n$ can be expressed in terms of linear and nonlinear modes, each being proportional to the same or lower order ``eccentricities'' and/or a combination of them. Constraints to hydrodynamic models can be imposed by measurements of the corresponding nonlinear response coefficients $\chi_{n(pk)}$, where $n$ represents the order of $V_n$ and $p$, $k$ with respect to lower-order symmetry plane angle or angles. Model calculations failed to simultaneously describe the measured $\chi$~\cite{Sirunyan:2019izh}, in particular, $\chi_{7223}$, as it can be seen in Fig.~\ref{fig:fig1} (right).

\section{Flow in proton-nucleus and proton-proton collisions}
\label{sec:SmallSystems}

\begin{figure}[!htb]
\centering
\begin{minipage}{0.5\textwidth}
\centering
\includegraphics[width=0.85\textwidth]{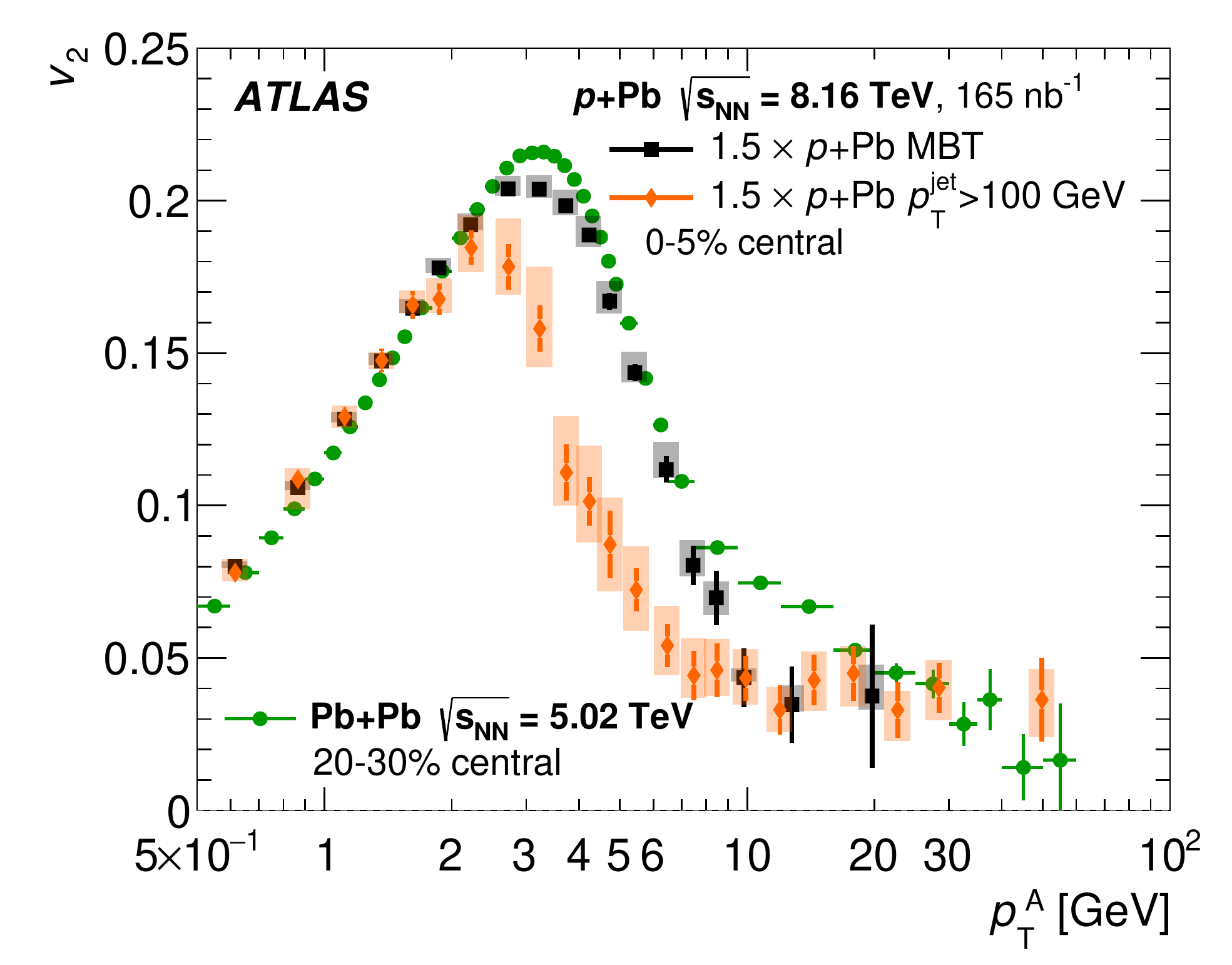}
\end{minipage}\hfill
\begin{minipage}{0.5\textwidth}
\centering
\includegraphics[width=0.8\textwidth]{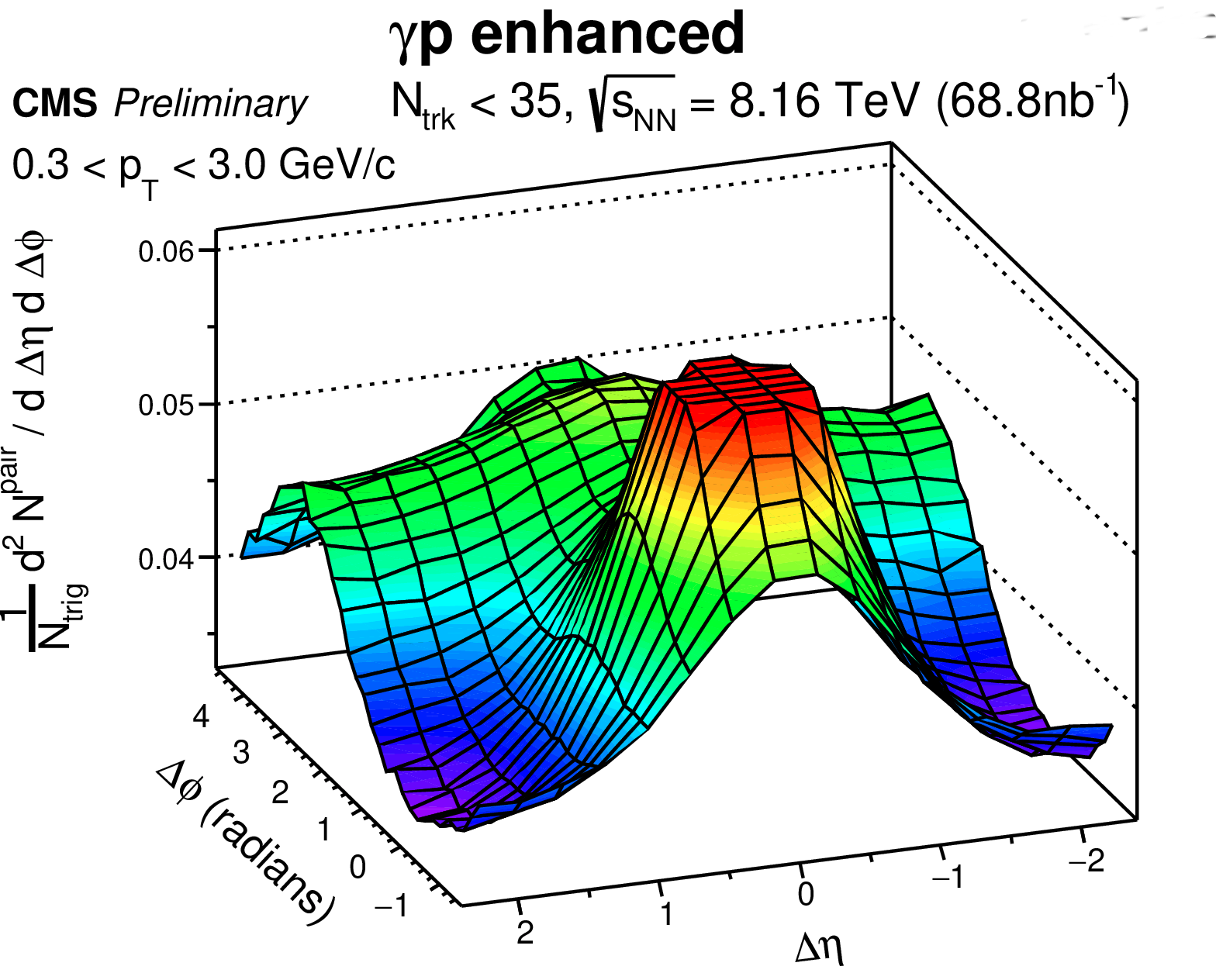}
\end{minipage}
\caption{Left: Distribution of $v_2$ as a function of the trigger-particle \pt, separately for minimum-bias and jet-enriched \Pp{}Pb events at 8.16\,\TeVns~\protect\cite{Aad:2019ajj}. Right: Two-dimensional correlation in photonuclear \Pp{}Pb events at 8.16\,\TeVns~\protect\cite{CMS:2020rfg}.}
\label{fig:fig2}
\end{figure}

A series of features from small systems are indicative of a collective behavior driven by the initial-state geometry and final-state effects: the near-side ridge in two-particle correlations, the \pt and event activity dependence of $v_n$, the mass ordering of $v_n$, multiparticle cumulants and their ratios. One of the main puzzles for the creation of a medium similar to that formed in nucleus-nucleus collisions is the absence of ``jet quenching'', \ie, no strong suppression as manifested by $R_{\Pp{}\mathrm{Pb}}$ measurements. Alternatively, measurements of $v_n$ at high \pt study the path length dependence of parton energy loss. Indeed, in Fig.~\ref{fig:fig2} (left) positive $v_2$ values are seen up to high \pt irrespective of the event content~\cite{Aad:2019ajj}. Significant $v_2$ values are also observed in photonuclear events~\cite{Aad:2021yhy,CMS:2020rfg} (Fig.~\ref{fig:fig2} (right)), with $v_2$ having a similar \pt and event activity dependence as the hadronic collision systems.

In proton-proton collisions, it is even less clear what mechanism gives rise to the observed finite azimuthal anisotropy. Until now, hints towards a final-state description were given, \eg, the indication of a mass ordering~\cite{Khachatryan:2016txc} with multi-particle angular correlations~\cite{Khachatryan:2016txc,Aaboud:2017acw} or that of the long-range correlations being only slightly affected when particles associated with hard or semihard processes in the event are removed~\cite{ATLAS:2020guf}.

\section{Identified particle flow in all systems}
\label{sec:FlowID}

\begin{figure}[!htb]
\centering
\begin{minipage}{0.5\textwidth}
\centering
\includegraphics[width=0.85\textwidth]{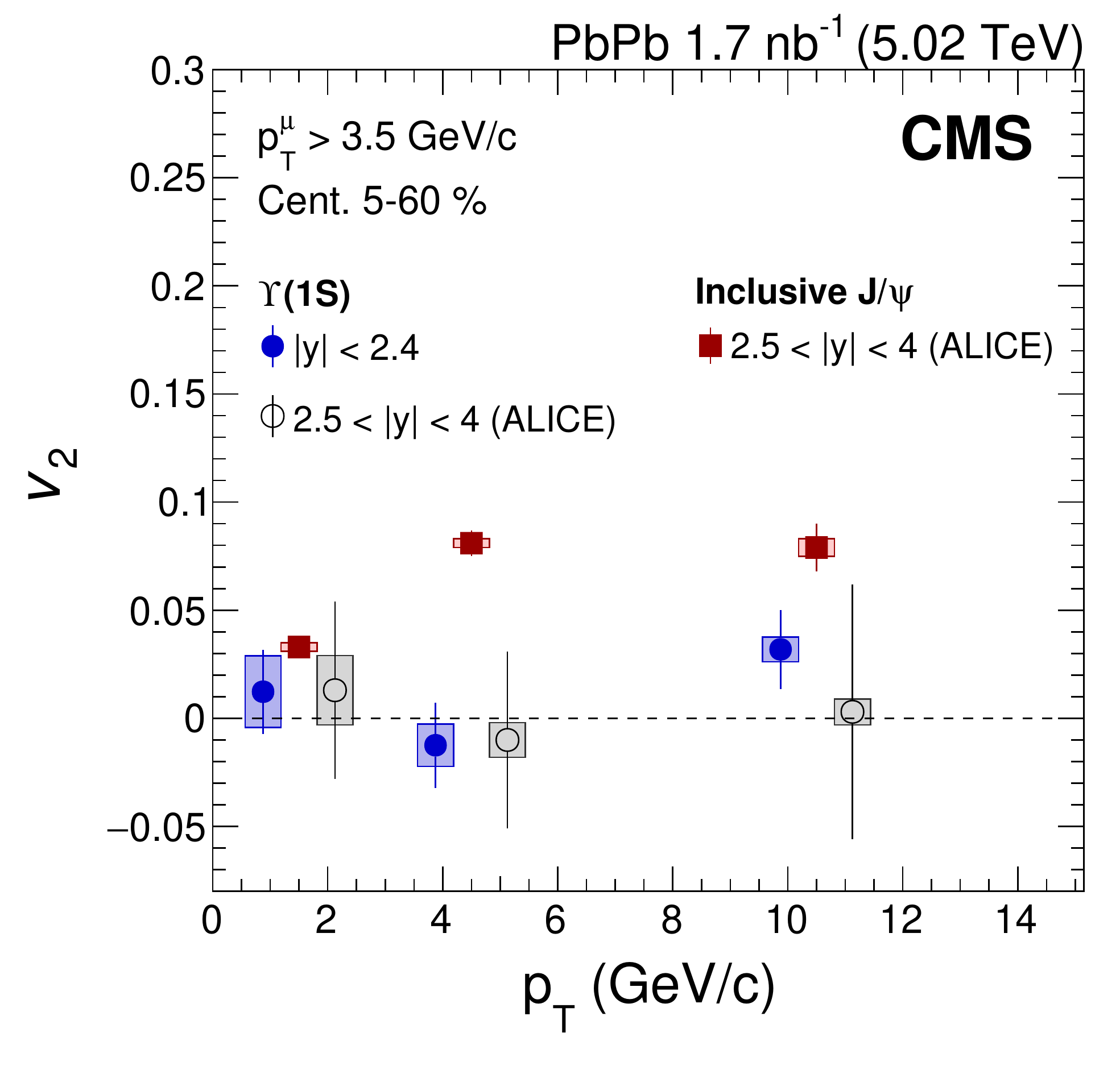}
\end{minipage}\hfill
\begin{minipage}{0.5\textwidth}
\centering
\includegraphics[width=0.99\textwidth]{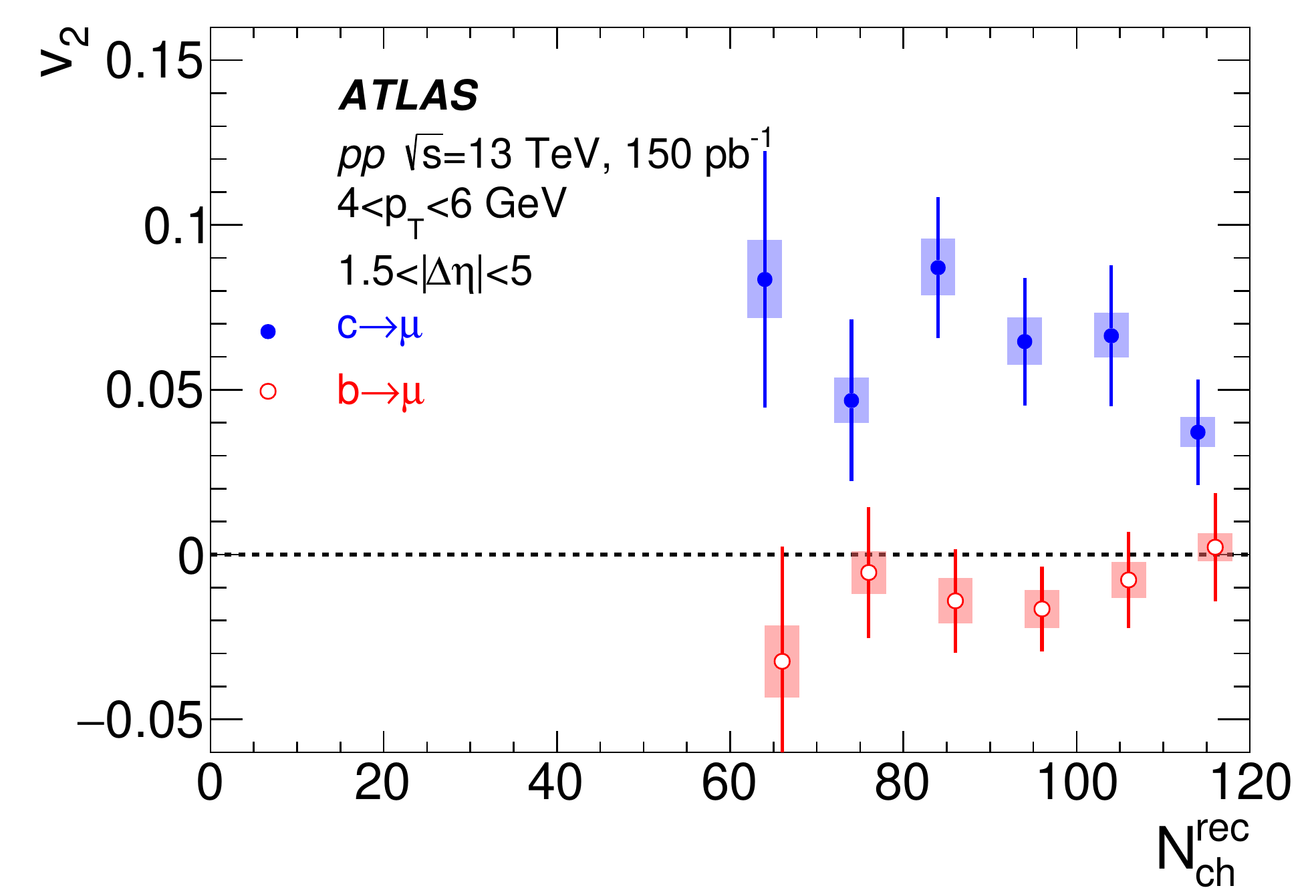}
\end{minipage}
\caption{Left: Transverse momentum dependence of $v_2$ of \PGU{}(1S) in PbPb collisions at 5.02\,\TeVns~\protect\cite{Sirunyan:2020qec}, compared to forward \PGU{}(1S) and \JPsi measurements from the ALICE Collaboration. Right: Elliptic flow $v_2$ of muons originating from charm or bottom hadrons as a function of the reconstructed track multiplicity in pp collisions at 13\,\TeVns~\protect\cite{Aad:2019aol}.}
\label{fig:fig3}
\end{figure}

Results of inclusive charged hadrons support the final-state scenario with a fluid being created in heavy ion and at least in proton-nucleus collisions. Measurements of identified hadrons confirm these observations while paving the way for stringent tests of theory predictions on the grounds that heavy flavor particles are formed early and subsequently participate in the medium evolution. Flow of quarkonia study the interaction of charm and beauty with the QGP. While \JPsi and, for the first time with a multiparticle long-range correlation technique~\cite{CMS:2021rlx}, \PDz exhibit a significant flow, the $v_2$ of \PGU{}(1S) and \PGU{}(2S)~\cite{Sirunyan:2020qec} is consistent with zero (Fig.~\ref{fig:fig3} (left)). Measurements of open heavy flavor hadrons also confirm a positive $v_2$, with the possibility to disentangle contributions from bottom hadrons, which are indeed found to flow less than charm hadrons~\cite{Aad:2020grf,Aad:2019aol,Sirunyan:2020obi}  (Fig.~\ref{fig:fig3} (right)). 

\section{Summary}
\label{sec:summary}

Although significant progress has been made on the level of precision achieved at large collision systems, the amount of data collected at LHC allow the measurement of more complex flow-related observables. More specifically, such measurements pose constraints on initial conditions, which in turn contribute to more precise modeling of the final-state dynamics. Most  previous  flow  measurements  focused on $V_n$ (overall  flow),  \ie, $v_n$ with respect to $\Psi_n$, but recently event-by-event flow fluctuations, longitudinal flow decorrelations, and measurements of nonlinear response coefficients demonstrate that model calculations cannot simultaneously describe all the aforementioned observables.

Multiparticle long-range correlations of inclusive charged and identified hadrons are observed down to the smallest collision systems. The origin of positive $v_n$ seen up to high \pt is however still not resolved. Whether this is a manifestation of a collective behavior of the system created in such collisions and/or a dilute system with parton scatterings requires experimental techniques for suppressing nonflow contamination. In parallel, finding ways to discriminate between the two different scenarios are desirable, \eg, new results provide information on the QCD dynamics of multiparticle production in high-energy photonuclear collisions.

\section*{Acknowledgments}

Supported by the Nuclear Physics program \href{https://pamspublic.science.energy.gov/WebPAMSExternal/Interface/Common/ViewPublicAbstract.aspx?rv=d1ddcae6-235b-4163-ae34-01fce58e5f90&rtc=24&PRoleId=10}{DE-SC0019389} of the U.S. Department of Energy.

\section*{References}

\bibliography{GKKrintiras}

\end{document}